\documentclass[aps,pra,onecolumn,superscriptaddress,12pt]{revtex4-1}
\pdfoutput=1

\usepackage[utf8]{inputenc}
\usepackage{amssymb,amsmath}
\usepackage{graphicx}
\usepackage[usenames,dvipsnames]{xcolor}
\usepackage[colorlinks,bookmarks=false,citecolor=blue,linkcolor=black,urlcolor=blue]{hyperref}
\usepackage{amssymb,amsmath}
\usepackage[vcentermath]{youngtab} 
\usepackage{subfig}

\newcommand{\SU}{\ensuremath{\mathrm{SU}}}
\newcommand{\su}{\ensuremath{\mathrm{su}}}
\newcommand{\U}{\ensuremath{\mathrm{U}}}

\newcommand{\kk}{\ensuremath{\mathcal{K}}}
\newcommand{\HH}{\ensuremath{\mathcal{H}}}
\newcommand{\idm}{\ensuremath{1\!\!1}}

\newcommand{\CC}{\mathbb{C}}
\newcommand{\C}[1]{C^{\scalebox{0.85}{#1}}}
\newcommand{\D}[1]{D^{\scalebox{0.85}{#1}}}
\newcommand{\rest}{\Big|_{\scalebox{0.85}{\su(2)}}}

\newcommand{\ts}{\textsuperscript}

\begin{document}

\title{The bilinear-biquadratic model on the complete graph}

\author{D\'avid Jakab}
\affiliation{Institute for Solid State Physics and Optics - Wigner Research Centre for Physics, Hungarian Academy of Sciences, H-1525 Budapest P.O. Box 49, Hungary}
\affiliation{Institute of Physics - University of Pécs, H-7624 Pécs, Ifjúság u.~6, Hungary}

\author{Gergely Szirmai}
\affiliation{Institute for Solid State Physics and Optics - Wigner Research Centre for Physics, Hungarian Academy of Sciences, H-1525 Budapest P.O. Box 49, Hungary}

\author{Zolt\'an Zimbor\'as}
\affiliation{Department of Theoretical Physics - Wigner Research Centre for Physics, Hungarian Academy of Sciences, H-1525 Budapest P.O. Box 49, Hungary}

\date{\today}

\begin{abstract}
We study the spin-$1$ bilinear-biquadratic model on the complete graph of $N$ sites, i.e., when each spin is interacting with every other spin with the same strength. Because of  its complete permutation invariance, this  Hamiltonian can be rewritten as the linear combination of the quadratic Casimir operators of \su(3) and \su(2). Using group representation theory, we explicitly diagonalize the Hamiltonian and map out the ground-state phase diagram of the model. Furthermore, the complete energy spectrum, with degeneracies, is obtained analytically for any number of sites. 
\end{abstract}

\maketitle

\section{Introduction}
\label{intro}

The bilinear-biquadratic (BLBQ) model is the generalization of the spin-1/2 Heisenberg model to spin-1 systems with the most general rotation invariant nearest-neighbor spin interaction \cite{papanicolaou1986ground,fath1991period}. Its Hamiltonian is given by
\begin{equation}
\label{eq:BLBQH}
H_{\mathrm{BLBQ}}=\sum_{\left<i,j\right>} \left[
\cos(\gamma) \left(\mathbf{S}_i\cdot\mathbf{S}_j\right) + 
\sin(\gamma) \left(\mathbf{S}_i\cdot\mathbf{S}_j\right)^2
\right],
\end{equation}
where $\mathbf{S}_i$ denotes the spin operator, and the sum is over neighboring sites $\left<i,j\right>$ of the underlying lattice. The parameter $\gamma$ determines the signs and relative strength of the bilinear and biquadratic terms. The energy is measured in units for which the squared sum of the prefactors before the two terms gives unity. As $s=1$, every monomial of $\mathbf{S}_i\cdot\mathbf{S}_j$ which is of higher order than the biquadratic term can be reexpressed as the linear combination of only the bilinear and the biquadratic terms (and the identity operator, which is usually dropped) \cite{fazekas1999lecture}. For $\gamma=0$, the usual spin-1 Heisenberg model with antiferromagnetic coupling is recovered, while for $\gamma=\pi$ the ferromagnetic case is realized. There are two other special points of the model: $\gamma=\pi/4$ and $\gamma=5\pi/4$. In these points $\sin\gamma=\cos\gamma$, and the symmetry group of the model is enlarged from the rotation group to \SU(3) \cite{penc2011spin}.

The first golden era of the BLBQ model was in the mid '80s, after Haldane's discovery that spin-1 Heisenberg chains can have a gapped excitation spectrum in contrary to spin-1/2 systems, where the spectrum is always gapless \cite{haldane1983continuum,haldane1983nonlinear}. This remarkable difference initiated an intensive study of the BLBQ chain, and its phase diagram is mostly understood by now. Exact Bethe ansatz solutions exist for $\gamma=-\pi/4$ \cite{takhtajan1982picture,babujian1982exact,babujian1983exact} and $\gamma=\pi/4$ \cite{uimin1970one,lai1974lattice,sutherland1975model}. When $\gamma=\arctan(1/3)$, Eq.~\eqref{eq:BLBQH} is just the parent Hamiltonian of the AKLT state \cite{affleck1987rigorous,affleck1988valence}. The AKLT state corresponds to a specific point inside the gapped Haldane phase \cite{haldane1983continuum,haldane1983nonlinear}, located between $-\pi/4<\gamma<\pi/4$, which is an example of symmetry protected topological phases \cite{pollmann2012symmetry}. The $\pi/4<\gamma<\pi/2$ region is an extended critical phase with strong antiferroquadrupolar correlations. For $\pi/2<\gamma<5\pi/4$ ferromagnetic correlations dominate the ground state. Finally, if $5\pi/4<\gamma<7\pi/4$ the system is in a dimerized regime, which is again a gapped phase. There is another, conjectured, critical phase between the ferromagnetic and the dimerized phases, which was proposed by Chubukov \cite{chubukov1990fluctuations,chubukov1991spontaneous}. However, there is a still ongoing debate about its existence \cite{rizzi2005phase,buchta2005probable,porras2006renormalization,lauchli2006spin,hu2014berry}. 

We know much less about the phases on two- and higher-dimensional lattices. Although in $D=2$ and higher, especially on bipartite lattices, symmetry-breaking states start to be more frequent, and mean-field theories are performing better. On the two-dimensional square lattice, mean-field theory predicts a conventional ferromagnetic state for $\pi/2<\gamma<5\pi/4$, a ferroquadrupolar phase for $5\pi/4<\gamma<3\pi/2$, an antiferromagnetic (N\'eel ordered) phase between $-\pi/2<\gamma<\pi/4$, and a semi-ordered phase for $\pi/4<\gamma<\pi/2$ \cite{papanicolaou1988unusual}. The ferroquadrupolar state is a symmetry-breaking state with nonvanishing quadrupole moment, however, with zero magnetization \cite{penc2011spin}. In the semi-ordered phase, the variational calculation gives a highly degenerate manifold of differently ordered states. Possibly, the inclusion of fluctuations selects one of these potential candidates, and in the end, the ground state becomes an ordered state. In the square lattice, Quantum Monte Carlo simulation is possible for $\pi<\gamma<2\pi$, which confirms the mean-field results in this parameter range \cite{harada2001loop}.

A further motivation for the studies of the BLBQ model stems from the recent and still not completely explored possibilities in ultracold atom experiments. Dilute gas samples of ultracold atoms on optical lattices can emulate magnetic systems with unprecedented control (see Ref. \cite{lewenstein2012ultracold} as a review). The main idea is to trap a multicomponent atomic gas cloud in an optical lattice and drive it into a Mott insulator state with exactly one particle per site. The multicomponent nature of the atoms comes from the simultaneous trapping of the $2F+1$ different magnetic sublevels ($m_F=-F,\ldots , F$) of a specific hyperfine state, where $F$ is the magnitude of the hyperfine spin. Although these atoms are charge neutral particles, they still interact with short-range scattering, the strength of which can be controlled in a wide range through the access of various scattering resonances \cite{bloch2008many}. With the help of Rydberg atoms \cite{saffman2010quantum,weimer2010rydberg,takei2016direct} or ion traps \cite{richerme2014non,jurcevic2014quasiparticle,cohen2015simulating,senko2015realization} even long-range, many-body interacting, and higher-spin systems can be realized. Recently, even the infinite-range interaction needed for the topology of the complete graph was proposed for \SU(n) symmetric magnets with the help of a highly anharmonic trap \cite{beverland2016realizing}. Here, the key idea was to use extended box-potential orbitals instead of the localized Wannier states of a Mott insulator.

There are several theoretical methods to address the problem of interacting spin-1 systems. The most natural and least resource consuming one is the semiclassical method \cite{papanicolaou1988unusual}, which is a suitable generalization of spin-wave theory to the spin-1 case. One can expect this method to work in situations where fluctuations can be neglected, such as higher dimensional, bipartite lattices. When quantum fluctuations are important, numeric tools become more and more necessary. Quantum Monte Carlo methods \cite{harada2001loop,nightingale1986gap,harada2002quadrupolar} allow for moderate system sizes, but can suffer from the infamous sign problem, and their use is limited to certain parameter values. Another efficient set of numeric methods for moderate system sizes, or even for the thermodynamic limit, is based on tensor network algorithms \cite{rizzi2005phase,buchta2005probable,porras2006renormalization,jordan2008classical,niesen2017emergent,rodriguez2011field,dechiara2011bilinear}. These algorithms perform exceptionally in 1D (and can also be used in higher dimensions) provided that the energy gap is large enough. The most resource consuming method, of course, is brute-force exact diagonalization. This method is the only one, that can be applied to all parameter values and lattice configurations, however, as the Hilbert space grows as $3^N$ with $N$ the number of sites, its use is limited to very small system sizes \cite{bernu1992signature,bernu1994exact}. Recently, a very efficient technique was introduced for $\SU(n)$ antiferromagnets, in order to restrict the calculation to the singlet sector of the model, which is usually much smaller than the full Hilbert space \cite{nataf2014exact}.

In this paper, we study the BLBQ model in a situation where exact solution is possible for an arbitrary number of spins. To this end, we generalize the model from the lattice to the complete graph of $N$ nodes. 
 Classical spin models on complete graphs, such as the Curie-Weiss \cite{kochmanski2013curie} or the Sherrington-Kirkpatrick \cite{sherrington1975solvable} models, play an important role in statistical mechanics. The reason being that these can be treated relatively easily and still describe general features of the corresponding model on high-dimensional lattices. Also different quantum spin (and also fermion and boson) models have been investigated on complete graphs \cite{vidal2004entanglement,osborne2006statistics, chayes2008phase}. Due to recent results on quantum de Finetti theorems \cite{christandl2007one,kraus2013ground,werner2016quantum,KrumnowEisertZimboras17}, these models can also be interpreted as mean-field approximations to regular lattice models. However, let us mention that the complete graph is the most frustrated possible graph. Therefore, the ground state in the antiferromagnetic parameter region of  the BLBQ model will be very much unlike a classical mean-field state.
 
 The results obtained for fully connected systems are also useful for types of cluster mean-field theories where the  ``cluster'' is  chosen to be the complete graph \cite{weiss1948the}. In such an approach, one partitions the lattice into small identical clusters, the interactions within a cluster are described exactly, while the interactions between clusters are treated in a mean-field way. Also, apart from its appealing theoretical formulation, the complete-graph model has a possible realization in ultracold atom experiments with long-range interactions and also possible applications in metrology in such a fashion as was proposed for the \SU(n) model in Ref. \cite{beverland2016realizing}.

In spite of the apparent simplicity of the BLBQ model on a complete graph, the diagonalization of the Hamiltonian is far from trivial. The main tool used is the representation theory of Lie groups, since, as we show in Sec.~\ref{sec:blbq-ham}, the Hamiltonian can be interpreted as a linear combination of Casimir operators. Using this observation, in Sec.~\ref{sec:eigenspacedecomp} we introduce the quantum numbers that uniquely label the eigenspaces of the Hamiltonian, then in Sec.~\ref{sec:restriction} we provide the possible joint values of these quantum numbers. This allows us to determine the phase diagram and the spectrum of the model in Sec.~\ref{Sec:PhaseDiag} and Sec.~\ref{Sec:Spectrum}, respectively. Finally, in Sec.~\ref{Sec:discussion} we give a brief summary and outlook. Some of the background material and technical details are moved to the Appendix.

\section{The bilinear-biquadratic Hamiltonian on the complete graph}
\label{sec:blbq-ham}

As discussed in the Introduction, we replace the lattice with the complete graph of $N$ sites. That is, the Hilbert space is $(\CC^3)^{\otimes N}$, and each site neighbors all other sites. The graph is not bipartite any longer, and we have $N(N-1)/2$ interaction bonds. In this case, as we will show, the Hamiltonian is mapped to a linear combination of the $N$-site quadratic Casimir operators of the Lie algebras $\su(2)$ and $\su(3)$,
\begin{equation}
\label{eq:CH}
H=\sin(\theta)\,\C{\su(3)}+\cos(\theta)\,\C{\su(2)}.
\end{equation}
The Casimir operators corresponding to the $N$-fold tensor product of the 3-dimensional irreducible representations of $\su(2)$ and $\su(3)$, respectively, are defined as
\begin{subequations}
\label{eqs:cas}
\begin{flalign}
\C{\su(2)}&=\sum_{\mu=1}^{3}\sum_{k,l}S^\mu_{k}\, S^\mu_{l}, \label{eq:su2cas}\\
\C{\su(3)}&=\sum_{\alpha=1}^{8}\sum_{k,l}\Lambda^\alpha_{k}\,\Lambda^\alpha_{l}, \label{eq:su3cas}
\end{flalign}
\end{subequations}
where $k$ and $l$ go over the $N$ sites. We choose a representation where the single-site $\su(2)$ and $\su(3)$ generators are given by the usual rotation and Gell-Mann matrices,
\begin{subequations}
\label{eqs:rot}	
\begin{flalign}
S^1&=\left(\begin{array}{c c c}
0 & 0 & 0\\
0 & 0 & -i\\
0 & i & 0
\end{array}\right),&
S^2&=\left(\begin{array}{c c c}
0 & 0 & i\\
0 & 0 & 0\\
-i & 0 & 0
\end{array}\right),&
S^3&=\left(\begin{array}{c c c}
0 & -i & 0\\
i & 0 & 0\\
0 & 0 & 0
\end{array}\right),\label{eq:spinops}
\end{flalign}
\begin{flalign}
\Lambda^1&=\left(\begin{array}{c c c}
0 & 1 & 0\\
1 & 0 & 0\\
0 & 0 & 0
\end{array}\right),&
\Lambda^2&=\left(\begin{array}{c c c}
0 & -i & 0\\
i & 0 & 0\\
0 & 0 & 0
\end{array}\right),&
\Lambda^3&=\left(\begin{array}{c c c}
1 & 0 & 0\\
0 & -1 & 0\\
0 & 0 & 0
\end{array}\right),&
\Lambda^4&=\left(\begin{array}{c c c}
0 & 0 & 1\\
0 & 0 & 0\\
1 & 0 & 0
\end{array}\right),\label{eq:lambdaops1}\\
\Lambda^5&=\left(\begin{array}{c c c}
0 & 0 & -i\\
0 & 0 & 0\\
i & 0 & 0
\end{array}\right),&
\Lambda^6&=\left(\begin{array}{c c c}
0 & 0 & 0\\
0 & 0 & 1\\
0 & 1 & 0
\end{array}\right),&
\Lambda^7&=\left(\begin{array}{c c c}
0 & 0 & 0\\
0 & 0 & -i\\
0 & i & 0
\end{array}\right),&
\Lambda^8&=\frac{\sqrt{3}}{3}\left(\begin{array}{c c c}
1 & 0 & 0\\
0 & 1 & 0\\
0 & 0 & -2
\end{array}\right).\label{eq:lambdaops2}
\end{flalign}
\end{subequations}

To show the connection between Eq. \eqref{eq:CH} and the BLBQ Hamiltonian on the complete graph, let us first expand the $N$-site Casimir operators.
\begin{equation}
\label{eq:su2casexp}
\C{\su(2)}=\sum_k \sum_{\mu=1}^{3} S^\mu_k\,S^\mu_k + 2\sum_{(k,l)} \sum_{\mu=1}^{3} S^\mu_k\,S^\mu_l=2\,N\,\idm +2\sum_{(k,l)} \sum_{\mu=1}^{3} S^\mu_k\,S^\mu_l,
\end{equation}
where $(k,l)$ means the bond between the sites $k$ and $l$ (with $k \ne l$), and the corresponding sum runs over all the $N(N-1)/2$ different bonds. A similar expression can be obtained for the $\su(3)$ case,
\begin{align}
\C{\su(3)}&=\sum_k \sum_{\alpha=1}^{8} \Lambda^\alpha_k\,\Lambda^\alpha_k + 2\sum_{(k,l)} \sum_{\alpha=1}^{8} \Lambda^\alpha_k\,\Lambda^\alpha_l \nonumber \\
&=-\frac{8}{3}\,N \,(N-3)\,\idm + 4\sum_{(k,l)} \left[\sum\limits_{\mu=1}^3S^\mu_k\,S^\mu_l+\left(\sum\limits_{\mu=1}^3S^\mu_k\,S^\mu_l\right)^2\right]. \label{eq:su3casexp}
\end{align}
See Appendix~\ref{sec:appCasimir} for the detailed derivation.

Combining Eqs. \eqref{eq:CH}, \eqref{eq:su2casexp} and \eqref{eq:su3casexp}, and dropping the term proportional to the identity matrix, we arrive to
\begin{equation}
\label{eq:hammod}
H=\sum_{(k,l)}\left\lbrace\left[4\sin(\theta)+2\cos(\theta)\right]\left(\mathbf{S}_k\cdot\mathbf{S}_l\right)+4\sin(\theta)\left(\mathbf{S}_k\cdot \mathbf{S}_l\right)^2\right\rbrace.
\end{equation}
Therefore, if we replace the underlying lattice of the BLBQ model, given by Eq.~\eqref{eq:BLBQH}, with a complete graph, it becomes equivalent to the model given by Eq.~\eqref{eq:CH} with the following relation between $\gamma$ and $\theta$:
\begin{equation}
\label{eq:psitheta}
\tan(\gamma)= \frac{2\tan(\theta)}{1 + 2\tan(\theta)}.
\end{equation}

\section{Eigenspace decomposition of the Hamiltonian}
\label{sec:eigenspacedecomp}

Due to the permutation invariance of the Hamiltonian, its eigenspace decomposition can be obtained in an abstract form  from representation theoretic considerations. In this section, we introduce the notations and provide the decomposition of the Hilbert space into the eigenspaces of the Hamiltonian. Some of the basic definitions and concepts are summarized in Appendix~\ref{SchurWeylApp}.

Let us first note, that from a representation theoretic point of view, \su(3) contains two non-equivalent classes of \su(2) subalgebras. The first one contains those \su(2) subalgebras, which are unitarily equivalent to the one generated by the spin operators $S^1,S^2,S^3$ in Eq.~\eqref{eq:spinops}. In the second class, the \su(2) subalgebras are unitarily equivalent to that generated by $\Lambda^1,\Lambda^2,\Lambda^3$ in Eq.~\eqref{eq:lambdaops1}. Restricting the defining representation of \su(3) to the \su(2) subalgebras belonging to the first class yields the spin-1 representation, while the restriction to the subalgebras of the second class results in the direct sum of the spin-0 and the spin-$\frac{1}{2}$ representations. In the rest of the paper, the $\su(2) \subset \su(3) $ embedding will always refer to the \su(2) subalgebras of the first class.

On the Hilbert space $(\CC^3)^{\otimes N}$ of the $N$-site model, the relevant representations of \su(2) and \su(3) are the $N$-fold direct products of the spin-1 \su(2) and the defining representation of \su(3), respectively. These $N$-fold direct products of  representations
can be decomposed into direct sums of irreducible representations (irreps) of the corresponding groups:
\begin{subequations}
  \label{eq:repproduct}
  \begin{align}
  \left(\D{\su(2)}_1\right)^{\otimes N}&\cong\bigoplus_s m^{}_s \, \D{\su(2)}_s,\label{eq:repproduct2}\\
  \left(\D{\su(3)}_{(1,0)}\right)^{\otimes N}&\cong \bigoplus_{\lambda} m^{}_{\lambda}\, \D{\su(3)}_{\lambda}.\label{eq:repproduct3}
  \end{align}
\end{subequations}
In this decomposition, $m_s$, $m_\lambda$ are the multiplicities of the $\D{\su(2)}_s$, $\D{\su(3)}_{\lambda}$ irreps. Here $s$ labels the spin, which in our set-up is always integer. The symbol $\lambda$ labels Young diagrams of at most 2 rows and $N-3i$ boxes, $i=0,1,\cdots,\lfloor N/3\rfloor$. Equivalently $(\lambda_1, \lambda_2)$ integer pairs satisfying the conditions $\lambda_1+\lambda_2=N-3i$, $i=0,1,\cdots,\lfloor N/3\rfloor$ and $\lambda_1 \ge \lambda_2 \ge 0$. This labeling is explained in more detail along with the Schur-Weyl duality in Appendix~\ref{SchurWeylApp}. In other words, the Hilbert space has the following two decompositions into a direct sum of irreducible subspaces $\HH_s$ and $\HH_\lambda$ 
\begin{equation}
\label{eq:decomp1}
  (\CC^3)^{\otimes N}=\bigoplus_s\left(\kk_s\otimes\HH_s\right)=\bigoplus_\lambda\left( \kk_\lambda\otimes\HH_\lambda \right),
\end{equation}
where the dimension of the multiplicity spaces $\kk_s$ and $\kk_\lambda$ is $m_s$ and $m_\lambda$, respectively. The value of $m_s$ can be calculated with the usual spin addition, while $m_\lambda$ is given by Eq.~\eqref{eq:hook}. As \su(2) is a subalgebra of \su(3), each \su(3) irreducible subspace in $(\CC^3)^{\otimes N}$ is a direct sum of \su(2) irreducible subspaces:
 \begin{equation}
   \label{eq:decomp}
  \D{\su(3)}_\lambda\rest\cong\bigoplus_{s}m^{(\lambda)}_{s}\D{\su(2)}_{s},\qquad \HH_\lambda=\bigoplus_{s}\left( \kk_{s}^{(\lambda)}\otimes\HH_{s}^{(\lambda)} \right).
\end{equation}
The left-hand side of the first equation denotes the restriction of the $\D{\su(3)}_\lambda$ irrep to the $\su(2)$ subalgebra of \su(3), the corresponding multiplicity spaces and irreducible subspaces are $ \kk_{ s}^{(\lambda)}$ and $\HH_{ s}^{(\lambda)}$, respectively. The dimension of $\kk_{ s}^{(\lambda)}$ is $m^{(\lambda)}_{s}$. The compatibility between the two decompositions of the Hilbert space in Eq. \eqref{eq:decomp} implies   $\sum_\lambda m^{(\lambda)}_{s}=m_{s}$, and $\sum_s m^{(\lambda)}_{s}=m_{\lambda}$.  Introducing the notation $s \prec \lambda$, which will mean that $\D{\su(2)}_s$ appears in the irreducible decomposition of $\D{\su(3)}_\lambda\rest$ (i.e. $m^{(\lambda)}_{s} \ne 0$), the complete $N$-particle Hilbert space can be written as the following direct sum of subspaces
\begin{equation}
  \label{eq:decomp2}
(\CC^3)^{\otimes N}= 
\bigoplus_\lambda \bigoplus_{s \prec \lambda} \kk_\lambda\otimes \kk^{(\lambda)}_s \otimes \HH^{(\lambda)}_s.
\end{equation}

The BLBQ Hamiltonian, Eq. \eqref{eq:CH}, on the complete graph is a linear combination of the \su(2) and \su(3) quadratic Casimir operators for $N$ sites. This implies that the subspace $\kk_\lambda\otimes \kk^{(\lambda)}_s \otimes \HH^{(\lambda)}_s$ is an eigenspace of $H$. Let $P^{(\lambda)}_s$ denote the projection to this subspace, i.e., to the spin-$s$ subspace of $\D{\su(3)}_{\lambda}\rest$. Now the Hamiltonian takes the form
\begin{subequations}
\label{eqs:Hamiltoniandiag}
\begin{align}
H&=\sum\limits_{\lambda}\sum\limits_{s\prec\lambda} E^{(\lambda)}_s\,P^{(\lambda)}_s, \label{Hamiltoniandiag}\\
E^{(\lambda)}_s&=\frac{4}{3}\sin(\theta)(\lambda_1^2+\lambda_2^2-\lambda_1\lambda_2+3\lambda_1)+\cos(\theta)s(s+1)\label{eq:eigenvalue},
\end{align}
\end{subequations}
where $E^{(\lambda)}_s$ is the eigenvalue corresponding to the subspace of $P^{(\lambda)}_s$. The first term is the value of the quadratic Casimir of \su(3) in the $(\lambda_1, \lambda_2)$ irrep \cite{iachello}, while the second term is the usual total spin squared, i.e., the Casimir of \su(2).

Finding the ground state of \eqref{Hamiltoniandiag} means finding the quantum numbers $\lambda_1,\lambda_2$ and $s$, for which the energy in Eq.~\eqref{eq:eigenvalue} is minimal. However, the value of $s$ cannot be chosen independently from those of $\lambda_1$ and $\lambda_2$, because not all spin-$s$ representations appear in the \su(3) irrep characterized by $\lambda_1$ and $\lambda_2$. This is why the knowledge of $m^{(\lambda)}_{s}$ in Eq.~\eqref{eq:decomp} is so important: only if $m_{\lambda}$ and $m^{(\lambda)}_{s}$ are nonzero, does an eigenspace with quantum numbers $\lambda_1$, $\lambda_2$, and $s$ appear
in the decomposition Eq.~\eqref{eq:decomp2}, and consequently in Eq.~\eqref{Hamiltoniandiag}.

\section{Restricting su(3) representations to the su(2) subalgebra}
\label{sec:restriction}

In this section, we calculate the multiplicities, $m_s^{(\lambda)}$, appearing in the decomposition \eqref{eq:decomp}. Restricting the trivial representation of $\su(3)$ simply yields the spin-0 irrep of $\su(2)$, i.e., $m_0^{(0,0)}=1$. The defining representation of $\su(3)$ maps directly to the spin-1 irrep, with $m_1^{(1,0)}=1$.
For the rest, we start by giving a general rule for the $\su(3)$ irreps characterized by the Young diagrams with only one row, i.e., we decompose
$\D{\su(3)}_{(\lambda_1,0)}\rest$. Then, by recursion, we obtain the rule for the other irreps.

The \su(3) irrep corresponding to the single-row Young diagram $(N,0)$ is supported on the completely symmetric part of the Hilbert space, $\mathcal{S}[(\CC^3)^{\otimes N}]$, with $\mathcal{S}$ being the operator projecting to the symmetric subspace. This subspace can be decomposed into a direct sum of \su(2) irreducible subspaces as seen in Eq. \eqref{eq:decomp}; and each spin-s \su(2) irreducible subspace can be decomposed into eigenspaces of the z-component of the total spin operator, $S^z=\sum\limits_kS_k^z$. We denote the subspace corresponding to eigenvalue $\ell\in\lbrace-s,\ldots,s\rbrace$ of the spin-s irrep by $V_{\ell}^{(s)}$, and the decomposition reads as
\begin{equation}
  \mathcal{S}[(\CC^3)^{\otimes N}]=\bigoplus_{s}\left( \kk_{s}^{(N,0)}\otimes\HH_{s}^{(N,0)} \right)=\bigoplus_s\left( \kk_s^{(N,0)}\otimes \bigoplus_{\ell=-s}^sV_\ell^{(s)}  \right).
  \label{eq:sajt}
\end{equation}
Let $V_\ell$ denote the  $S^z$ eigenspace with eigenvalue $\ell$ in the symmetric subspace.
The eigenvalue $\ell$ can only appear in spin-$s$ representations with $s \ge \ell$, thus
\begin{equation}
V_\ell= \bigoplus_{s\ge|\ell|}\left(  \kk_s^{(N,0)}\otimes V_\ell^{(s)}\right).
\end{equation}
Furthermore, since every $V_\ell^{(s)}$ is one-dimensional, the dimension of $V_\ell$ is equal to the sum of multiplicities of spin-s irreps with $s\ge|\ell|$:
\begin{equation}
  \text{dim}\left[ V_\ell \right]=\sum\limits_{s\ge|\ell|}m^{(N,0)}_s.
\end{equation}
Thus, we proceed by figuring out these dimensions.

The $\ell=N$ eigenvalue can only come from the $|1,1,\cdots,1\rangle$ vector, hence $m^{(N,0)}_N=1$. A basis of the $\ell=N-1$ eigenspace in $(\CC^3)^{\otimes N}$ can be constructed from the previous vector by lowering one of the spins to 0. This subspace is N-dimensional, but its intersection with the symmetric part is only 1-dimensional spanned by the vector $|0,1,1,\ldots,1\rangle+|1,0,1,\ldots,1\rangle+ \cdots +|1,1,1,\ldots,0\rangle$. This implies $ \text{dim}\left[ V_{N-1} \right]=1$ and $m^{(N,0)}_{N-1}=0$.
We can get a basis of the $\ell=N-2$ eigenspace of $(\CC^3)^{\otimes N}$ from the  $|1,1,\cdots,1\rangle$ vector by either lowering two spins to 0, or by lowering one to -1. Any of the basis elements with 0 spin on two sites have the same projection to the symmetric subspace, and similarly only one symmetric vector can be generated from the basis elements with a single -1 spin, so $\text{dim}(V_{N-2})=2$ and $m^{(N,0)}_{N-2}=1$.

Generalizing this procedure, we can see that for an arbitrary $\ell$ we can lower $a$ spins to $0$ and $b$ spins to -1 such that $\ell=N-(a+2b)$. Each different choice of $a$ and $b$ corresponds to one linearly independent element of $V_\ell$. Therefore, the number of integer solutions for $a,b \ge 0$ of the previous equation is the dimension of $V_\ell$, resulting in  
\begin{equation}
  \text{dim}(V_\ell) = \lfloor (N-|\ell|)/2\rfloor +1.
\end{equation}
Here, $\lfloor x \rfloor$ denotes the floor function, i.e., the largest integer not bigger than x. As the dimension of the $S^{z}$ eigenspace changes by 1 at every second $\ell$, we can infer that only every second spin-s irrep is present in the decomposition \eqref{eq:decomp} of the completely symmetric subspace of $(\CC^3)^{\otimes N}$, and with a multiplicity of 1.
Therefore,
\begin{equation}
\D{\su(3)}_{(\lambda_1,0)}\rest= 
\begin{cases}
\bigoplus_{s=0}^{\lfloor \lambda_1/2\rfloor} \D{\su(2)}_{2s+1} &  \text{if $\lambda_1$ is odd,}\\[2mm]
\bigoplus_{s=0}^{\lambda_1/2} \; \; \D{\su(2)}_{2s} &  \text{if $\lambda_1$ is even.}
\end{cases}
\end{equation}
In terms of multiplicities this means 
\begin{equation}
  \label{eq:mult0}
  m_s^{(\lambda_1,0)}=
  \begin{cases}
    1-\mathrm{mod}(s+\lambda_1,2) & \text{if } 0\le s\le \lambda_1,\\
    0 & \text{otherwise},
  \end{cases}
\end{equation}
where $\mathrm{mod}(n,2)$ is the reminder after the division of $n$ by 2.
\begin{figure}[bt!]
  \centering
  \includegraphics[scale=1.4]{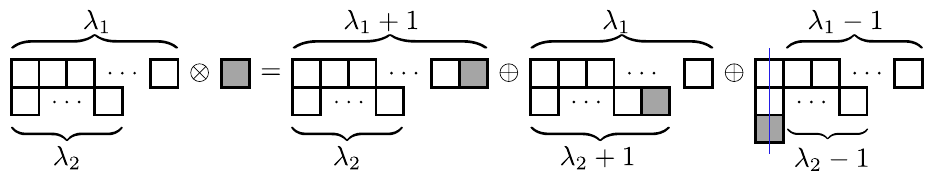}
  \caption{Illustration for decomposing the product of the $(\lambda_1,\lambda_2)$ and $(1,0)$ representations into the direct sum of $\su(3)$ irreps.}
  \label{fig:multrule1}
\end{figure}

\begin{figure}[tb!]
   \centering
   \includegraphics[scale=1.4]{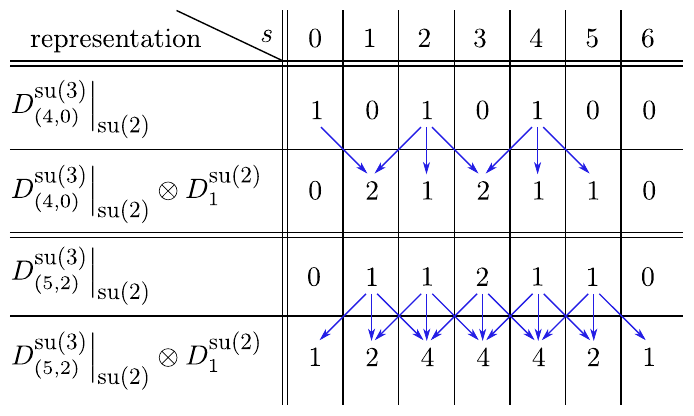}
   \caption{Illustration of the calculation of the multiplicities in the decomposition Eq.~\eqref{eq:lhsmultruledec}.}
   \label{fig:multrule2}
 \end{figure}

Before continuing with the more complicated cases, we first introduce the rule for directly multiplying an arbitrary $\su(3)$ irrep with the defining representation \cite{iachello}: The resulting representation is the direct sum of all possible diagrams we can get by attaching one box to the original diagram. We also need to take into account the equivalence relation between $\su(3)$ diagrams described in Appendix~\ref{SchurWeylApp}, that is, we are free to delete any columns of height 3 from the left of each diagram. The multiplication rule,  illustrated in Fig.~\ref{fig:multrule1}, reads as
\begin{equation}
  \label{eq:multrule}
  \D{\su(3)}_{(\lambda_1,\lambda_2)}\otimes \D{\su(3)}_{(1,0)}=\D{\su(3)}_{(\lambda_1+1,\lambda_2)}\oplus \D{\su(3)}_{(\lambda_1,\lambda_2+1)}\oplus \D{\su(3)}_{(\lambda_1-1,\lambda_2-1)}. 
\end{equation}
Let us restrict both sides of the equation to $\su(2)$. The left-hand side of this equation decomposes into $\su(2)$ irreps using the usual multiplication rule for spins,
\begin{equation}
\D{\su(2)}_j\otimes \D{\su(2)}_1=\bigoplus\limits_{s=|j-1|}^{j+1}\D{\su(2)}_s.
\end{equation}
Furthermore, using Eq. \eqref{eq:decomp},
\begin{multline}
\label{eq:lhsmultruledec}
\D{\su(3)}_{(\lambda_1,\lambda_2)}\rest\otimes \D{\su(3)}_{(1,0)}\rest\cong\bigoplus\limits_{s=0}^{\infty} m_s^{(\lambda_1,\lambda_2)} \D{\su(2)}_s\otimes \D{\su(2)}_1\\
=m_1^{(\lambda_1,\lambda_2)}\D{\su(2)}_0\oplus \bigoplus\limits_{s=1}^{\infty} \Big[ m_{s-1}^{(\lambda_1,\lambda_2)}+m_{s}^{(\lambda_1,\lambda_2)}+m_{s+1}^{(\lambda_1,\lambda_2)}
\Big] \D{\su(2)}_s.
\end{multline}
This step is illustrated in Fig.~\ref{fig:multrule2}. The right-hand side of the restriction of Eq. \eqref{eq:multrule} becomes
\begin{multline}
  \label{eq:rhsmultruledec}
  \D{\su(3)}_{(\lambda_1+1,\lambda_2)}\rest\oplus \D{\su(3)}_{(\lambda_1,\lambda_2+1)}\rest\oplus \D{\su(3)}_{(\lambda_1-1,\lambda_2-1)}\rest\\
\cong\bigoplus\limits_{s=0}^{\infty}
\Big[ m_{s}^{(\lambda_1+1,\lambda_2)} + m_{s}^{(\lambda_1,\lambda_2+1)} + m_{s}^{(\lambda_1-1,\lambda_2-1)}\Big] \D{\su(2)}_s.
\end{multline}
Combining Eqs.~\eqref{eq:lhsmultruledec}, \eqref{eq:rhsmultruledec} and the $\su(2)$ restriction of Eq.~\eqref{eq:multrule},
\begin{equation}
\label{eq:recurrenceeq}
(1-\delta_{s,0})\Big(m_{s-1}^{(\lambda_1,\lambda_2)}+m_s^{(\lambda_1,\lambda_2)}\Big)+m_{s+1}^{(\lambda_1,\lambda_2)}
=m_s^{(\lambda_1+1,\lambda_2)}+m_s^{(\lambda_1,\lambda_2+1)}+m_{s}^{\lambda_1-1,\lambda_2-1}.
\end{equation}
This equation can be extended to the $\lambda_1=0$ case by setting $m_s^{(\lambda_1, -1)}=0$.

The recurrence relation \eqref{eq:recurrenceeq}, together with Eq.~\eqref{eq:mult0} as initial condition, uniquely determines the multiplicities $m_s^{(\lambda_1, \lambda_2)}$. One can show by, e.g., by direct substitution, that the solution is the following:
\begin{subequations}
\label{eqs:reqsols}
  \begin{equation}
    \label{eq:reqsol}
    m^{(\lambda_1,\lambda_2)}_s = \mathrm{max}\left(  0,\mathrm{min}\left[m^{(\lambda_1,\lambda_2)}_{1,s}, m^{(\lambda_1,\lambda_2)}_{2,s}\right]\right), 
  \end{equation}
     with
\begin{flalign}
     m^{(\lambda_1,\lambda_2)}_{1,s} &= \bigg\lceil \frac{\lambda_1{-}|2s{-}\lambda_1-1|}{4}+\mathrm{mod}\bigg[(\lambda_1{+}1)\bigg(\frac{|2\lambda_2{-}\lambda_1|{+}\lambda_1{+}2}{2}\bigg), 2\bigg] \bigg[\mathrm{mod}(s+1,2)-\frac{1}{2}\bigg]\bigg\rceil,  \\
     m^{(\lambda_1,\lambda_2)}_{2,s} &= \bigg\lceil
\frac{\lambda_1-|2\lambda_2-\lambda_1|-1}{4}\bigg\rceil
       +\mathrm{mod}\bigg(\frac{|2\lambda_2-\lambda_1|+3\lambda_1 +2}{2},2\bigg)(s+\lambda_1+1).
  \end{flalign}
\end{subequations}
According to Eq.~\eqref{eq:decomp}, the explicit knowledge of the multiplicities \eqref{eqs:reqsols} means that we have obtained the direct sum expansion of the \su(3) irreps into \su(2) irreps. In the following, we will see how it can be used to diagonalize the Hamiltonian \eqref{eq:CH}.

\section{Phase diagram of the model}
\label{Sec:PhaseDiag}

\begin{figure}[tb!]
  \centering
  \includegraphics[scale=1.4]{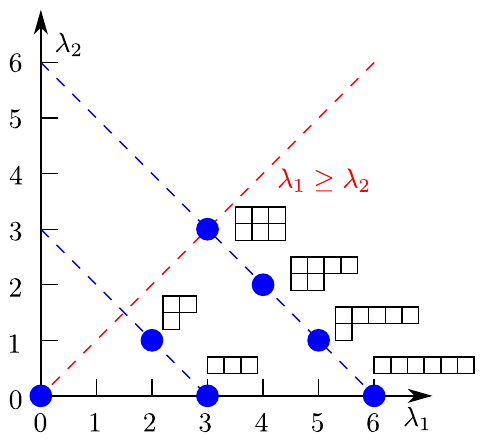} 
  \caption{The $(\lambda_1, \lambda_2)$ pairs (blue circles) corresponding to the $\su(3)$ irreps with nonzero multiplicity appearing in the decomposition Eq. \eqref{eq:repproduct3} for $N=6$.}
  \label{fig:possiblel1l2}
\end{figure}

In order to obtain the ground-state energy of the Hamiltonian \eqref{eq:CH}, we need to find the compatible  $\lambda_1$, $\lambda_2$ and $s$ quantum numbers that minimize its eigenvalue \eqref{eq:eigenvalue}. To describe these compatibility conditions, let us first note that $(\CC^3)^{\otimes N}$ factorizes into \su(3) irreducible subspaces according to Eq.~\eqref{eq:decomp1}.
Hence, the possible $\lambda=(\lambda_1,\lambda_2)$ pairs are the ones that appear with nonzero $m_\lambda$ multiplicities in Eq.~\eqref{eq:repproduct3}. Furthermore, for each such $(\lambda_1,\lambda_2)$ pair, the compatible $s$ values are those with $ m^{(\lambda_1,\lambda_2)}_s \ne 0$ in Eq.~\eqref{eq:decomp}. Given the optimal $\lambda=(\lambda_1, \lambda_2)$ and $s$ values, the ground-state eigenspace of the Hamiltonian is $\kk_\lambda\otimes \kk^{(\lambda)}_s \otimes \HH^{(\lambda)}_s$ from Eq.~\eqref{eq:decomp2}, which is usually degenerate. 

The solution of the ground-state energy problem is the simplest when the number of spins is divisible by $6$. Thus, we consider this situation first and explain the $\text{mod}(N,6)\ne 0$ case later. The divisibility by 3 is important for the following reason. According to Appendix~\ref{SchurWeylApp}, the $(\lambda_1,\lambda_2)$ irreducible subspaces present in the decomposition of the Hilbert space  $(\CC^3)^{\otimes N}$ correspond to Young diagrams with at most 2 rows and $N-3i$ boxes with $i=0,1,\ldots, {N/3} $ (see Fig.~\ref{fig:possiblel1l2}, for an illustration of the $N=6$ case). Consequently, the $(0, 0)$ diagram, or in other words, the \su(3) singlet is always present. Similarly, the divisibility by 2 is important to have the \su(2) singlet ($s=0$) included in the symmetric subspace.

\begin{figure}[tb!]
   \centering
   \subfloat[]{\includegraphics[height=0.38\linewidth]{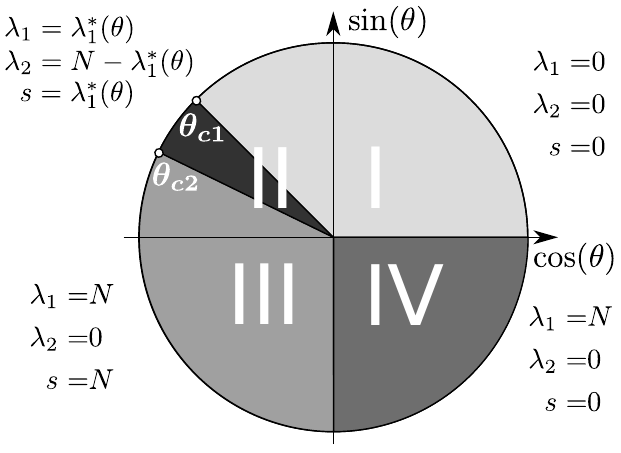}}
   \subfloat[]{\includegraphics[height=0.38\linewidth]{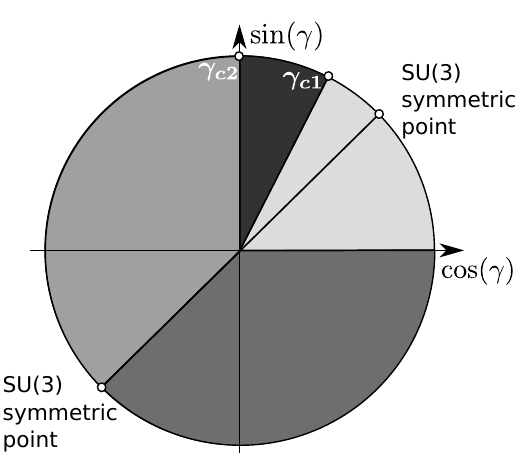}}
   \caption{The ground-state phase diagram of the BLBQ model on the complete graph. In subfigure (a) the phase diagram is plotted with the $\theta$ parameter used in Eq.~\eqref{eq:CH}. The quantum numbers are plotted next to the phase they belong to. In subfigure (b) we used the mapping Eq.~\eqref{eq:psitheta} to represent the phase diagram with the $\gamma$ parameter of the bilinear-biquadratic interaction.}
   \label{fig:phdiag}
 \end{figure}

The signs of the sine and the cosine prefactors are important in deciding the nature of the ground state, so we divide the phase diagram into four quarters using the angle $\theta$, see Fig.~\ref{fig:phdiag}(a). To minimize the second term of the energy \eqref{eq:eigenvalue} in the second and third quarters with a given $(\lambda_1, \lambda_2)$, where $\cos\theta<0$, we need to find the maximum possible $s$ value, while in the first and fourth quarters, where $\cos\theta>0$, we need to find the smallest allowed $s$. According to Eq.~\eqref{eqs:reqsols},
\begin{subequations}
\begin{flalign}
  \text{max}\{s|m^{\lambda_1,\lambda_2}_s\neq 0\}&=\lambda_1, \label{eq:maxs}\\
  \text{min}\{s|m^{\lambda_1,\lambda_2}_s\neq 0\}&=\begin{cases}
    0\quad\text{if both $\lambda_1$ and $\lambda_2$ are even,}\\
    1\quad\text{otherwise.}
    \label{eq:importantpart}
    \end{cases}
\end{flalign}
\end{subequations}
This means that in the second and third quarters we need to substitute $s=\lambda_1$ into Eq. \eqref{eq:eigenvalue}, then minimize the resulting expression:
\begin{equation}
  \label{eq:poly3rd}
E^{(\lambda)}_{\lambda_1} = (\frac{4}{3}\sin\theta+\cos\theta)\lambda_1^2 + (4\sin\theta+\cos\theta)\lambda_1 + \frac{4}{3}\sin\theta\lambda_2(\lambda_2-\lambda_1).
\end{equation}
On the other hand, in the first and fourth quarters, $s=0$ or $s=1$ has to be used according to Eq.~\eqref{eq:importantpart}. We will see shortly, that in the present case of $N$ being the multiple of $6$, both $\lambda_1$ and $\lambda_2$ are even in the ground state, so we use $s=0$ in Eq.~\eqref{eq:eigenvalue}, and finally arrive to the polynomial
\begin{equation}
  \label{eq:poly4th}
E^{(\lambda)}_{0} = \frac{4}{3}\sin\theta (\lambda_1^2 + \lambda_2^2 - \lambda_1\lambda_2+3\lambda_1).
\end{equation}
Let us now describe what happens in each of these quarters one by one.

\emph{In the first quarter} $(0<\theta<\pi/2)$, both the sine and the cosine prefactors are positive in Eq. \eqref{eq:eigenvalue} and $s=0$ can be taken. Hence, we need to minimize Eq.~\eqref{eq:poly4th}. Its minimum belongs to the $\su(3)$ singlet $(\lambda_1=0,\lambda_2=0)$, which appears since $N$ is divisible by 3. Therefore, the ground state in the whole region is both an \su(3) and an \su(2) singlet. Obviously, any \su(3) singlet is also an \su(2) singlet, thus the ground-state subspace is plainly the entire \su(3)-singlet subspace.
Using the hook length formula, Eq.~\eqref{eq:hook}, the dimension of the ground-state subspace is
\begin{equation}
 \text{dim}(\text{GS})=\frac{2N!}{(N/3+2)(N/3+1)^2(N/3!)^3}.
\end{equation}

\emph{In the fourth quarter} $(3\pi/2<\theta<2\pi)$, $\sin(\theta)$ is negative but $\cos(\theta)$ is positive. Now Eq.~\eqref{eq:poly4th} is minimized by the irrep $(\lambda_1,\lambda_2)=(N,0)$. Since $N$ is even, our earlier assumption that the ground state is in the $s=0$ subspace is justified. The single-row, N-box diagram labels the trivial representation of the permutation group, which is supported on the symmetric subspace of  $(\CC^3)^{\otimes N}$. According to Eq.~\eqref{eqs:reqsols}, $m^{(N,0)}_0=1$, i.e., there is only one \su(2) singlet in the symmetric subspace, implying that the ground state is non-degenerate, $\text{dim}(\text{GS})=1$.

\emph{In the third quarter} $(\pi<\theta<3\pi/2)$, both the sine and the cosine are negative, and Eq.~\eqref{eq:poly3rd} has to be minimized. Its minimum is at $(\lambda_1, \lambda_2)=(N,0)$. The symmetric subspace naturally contains the maximum spin ($s=N$) representation, $m^{(N,0)}_N=1$, which hence spans the ground-state subspace, yielding $\text{dim}(\text{GS})=2N+1$.

\emph{In the second quarter} $(\pi/2<\theta<\pi)$, the two terms in Eq. \eqref{eq:eigenvalue} are in competition with each other with different signs. In this case, the ground state is polarized according to Eq.~\eqref{eq:maxs}, i.e., $s=\lambda_1$. Again, we need to minimize Eq.~\eqref{eq:poly3rd}. Interestingly, the quantum numbers characterizing the ground state in the neighboring quarters extend quite a bit into the second quarter. The $\lambda_1=\lambda_2=s=0$ singlet phase is still the ground state for $\pi/2<\theta<\theta_{c1}$, while for $\theta_{c2}<\theta<\pi$ the ground state is in the subspace with $\lambda_1=s=N$, $\lambda_2=0$ as in the third quarter. In other words, the two phase boundaries move from $\pi/2$ and $\pi$ to $\theta_{c1}$ and $\theta_{c2}$. The phase boundary $\theta_{c1}$ is a complicated function of $N$. However, when $N$ approaches infinity, it is approximated by
\begin{equation}
  \label{eq:thetac1}
  \theta_{c1}=3\pi/4
\end{equation}
The other phase boundary, for any value of $N$, is given by
\begin{equation}
  \label{eq:thetac2}
  \theta_{c2}=\pi-\text{arctan}((2N+1)/(4N+4)).
\end{equation}

Between $\theta_{c1}$ and $\theta_{c2}$, the quantum numbers gradually change between those of the two neighboring phases. If $(\lambda_1,\lambda_2)$ were allowed to vary continuously, the pair that minimizes the energy would move on the $(x(\theta),N-x(\theta))$ line with 
  \begin{equation}
   x(\theta)=\frac{(4N-4)\sin(\theta)-\cos(\theta)}{8\sin(\theta)+2\cos(\theta)}. 
  \end{equation}
  Since only integer pairs are allowed, and the value of the energy \eqref{Hamiltoniandiag} on the $\lambda_1+\lambda_2=N$ line as a function of $\lambda_1$ is a positive parabola with its minimum at $x(\theta)$. The value of $\lambda_1$ in the ground state will be the closest integer value to $x(\theta)$, which is $\lambda_1^*(\theta)=\lceil {x(\theta)-1/2} \rceil$. This means that the optimal quantum numbers are $(\lambda_1, \lambda_2)=(\lambda_1^*(\theta), N-\lambda_1^*(\theta))$ and $s=\lambda_1^*(\theta)$. The dimension of the ground state subspace is
  \begin{equation} \text{dim}(\text{GS})=\frac{N!(2\lambda_1^*(\theta)-N+1)!}{(N-\lambda_1^*(\theta))!(\lambda_1^*(\theta)+1)!}(2\lambda_1^*(\theta)+1).
  \end{equation}

In summary, we have found an \su(3) singlet phase, in the region $0\le\theta\le\theta_{c1}$, a partially magnetized phase between $\theta_{c1}\le\theta\le\theta_{c2}$, a ferromagnetic phase between $\theta_{c2}\le\theta\le3\pi/2$, and a symmetric \su(2) singlet phase between $3\pi/2\le\theta\le2\pi$. The complete phase diagram is illustrated in Fig.~\ref{fig:phdiag}, both in terms of $\theta$, and in terms of the original parameter $\gamma$ used in Eq.~\eqref{eq:BLBQH}. In terms of the $\gamma$ angle, when $N$ is asymptotically large, the phase-transition points are at $\gamma=0, \arctan(2) ,\pi/2, 5\pi/4$. It is worth to note that the boundaries of the ferromagnetic phase are at the same $\gamma$ angles than for the one-dimensional BLBQ chain. 

It is also worth to discuss some special points of the phase diagram. When $\theta= \pi/2, 3\pi/2$  the Hamiltonian is $\SU(3)$ invariant.  The $\theta=3\pi/2$ point is at a phase boundary separating two phases within the symmetric subspace: the ferromagnetic and the symmetric singlet phases. At this point there is an even bigger degeneracy than in the two neighboring phases, all symmetric states are ground states. On the contrary, the $\SU(3)$ symmetry at $\theta=\pi/2$ does not result in additional degeneracy of the ground state. This is because the ground state in this case is an $\su(3)$ singlet, which means it also must be an $\su(2)$ singlet. Consequently, this phase extends over the $\theta=\pi/2$ point into the second quarter. The situation is similar with the $\theta=0,\pi$ points, where the Hamiltonian \eqref{eq:CH} contains only the \su(2) Casimir operator. On the one hand, $\theta=0$ separates the two \su(2) singlet phases, and at this point all \su(2) singlet states are ground states. On the other hand, for $\theta=\pi$ we have an \su(2) ferromagnetic state with maximal spin, which  is always contained in the symmetric subspace, thus, it does not result in  an additional degeneracy and does not separate phases.

 Finally, let us turn to the case when the number of spins is not divisible by $6$.  It is then possible that the Hilbert space does not have an $\su(3)$ singlet subspace, or that the $(N, 0)$ Young diagram has no $\su(2)$ singlet subspace.  In these cases  one cannot  separately optimize  the $\C{\su(2)}$ and $\C{\su(3)}$ parts of the Hamiltonian  in the second and the fourth quarters. Instead, there is a competition between two different ground states with almost optimal \su(3) and \su(2) quantum numbers.  For example, in the first quarter when $\text{mod}(N,3)=1$, the ground state for $0<\theta<\text{arctan}(1/3)$ has the quantum numbers $\lambda_1=2$, $\lambda_2=2$, $s=0$. When $\text{arctan}(1/3)<\theta<\pi/2$ this is replaced by the state with $\lambda_1=1$, $\lambda_2=0$, $s=1$. The difference in the quantum numbers is of the order of 1, and thus it is unimportant for large $N$.

 \section{Energy Spectrum}
 \label{Sec:Spectrum}
 
 \begin{figure}[tb!]
   \centering
   \subfloat[]{\includegraphics[width=0.5\linewidth]{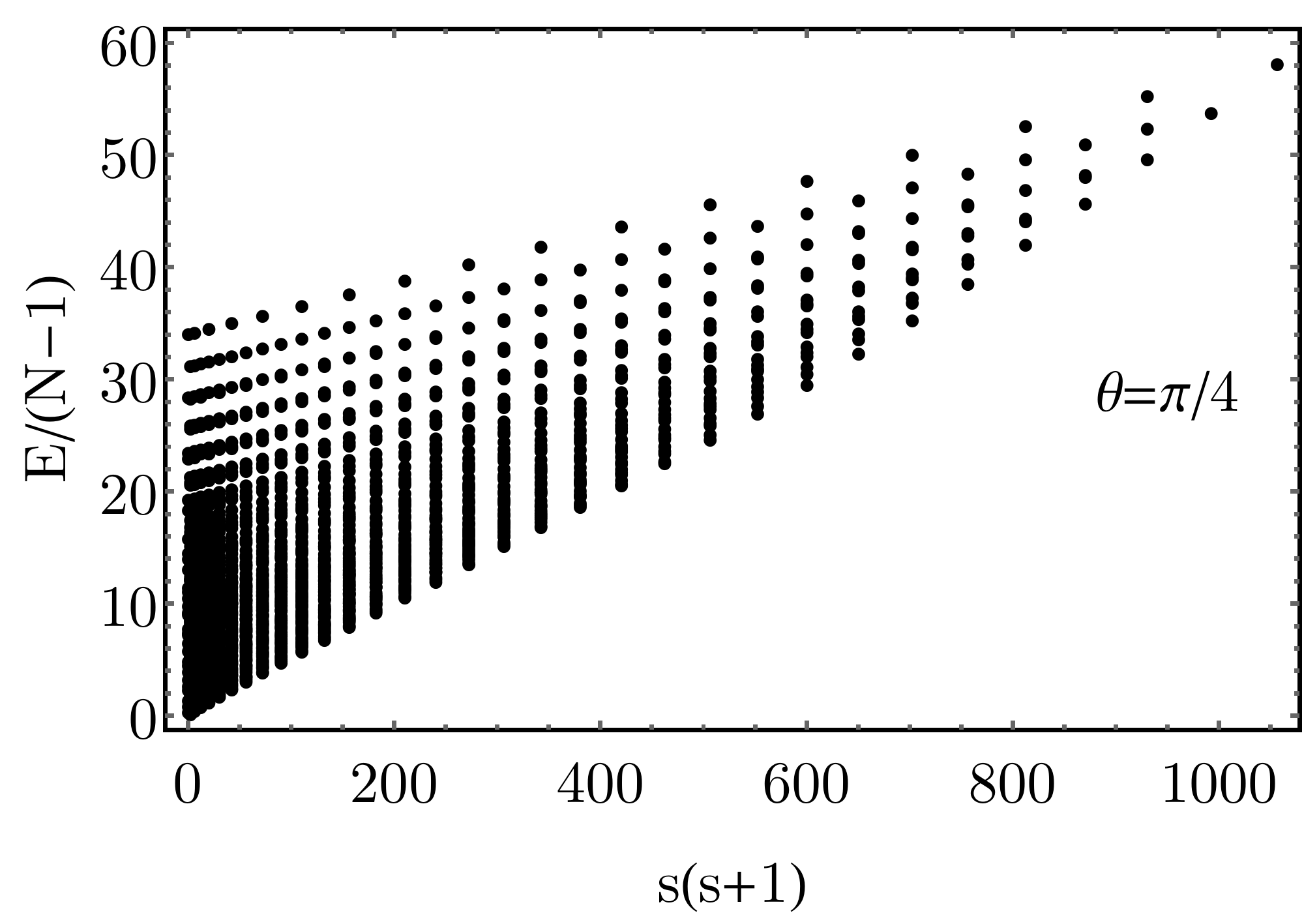}
   \label{fig:sp1}}
   \subfloat[]{\includegraphics[width=0.5\linewidth]{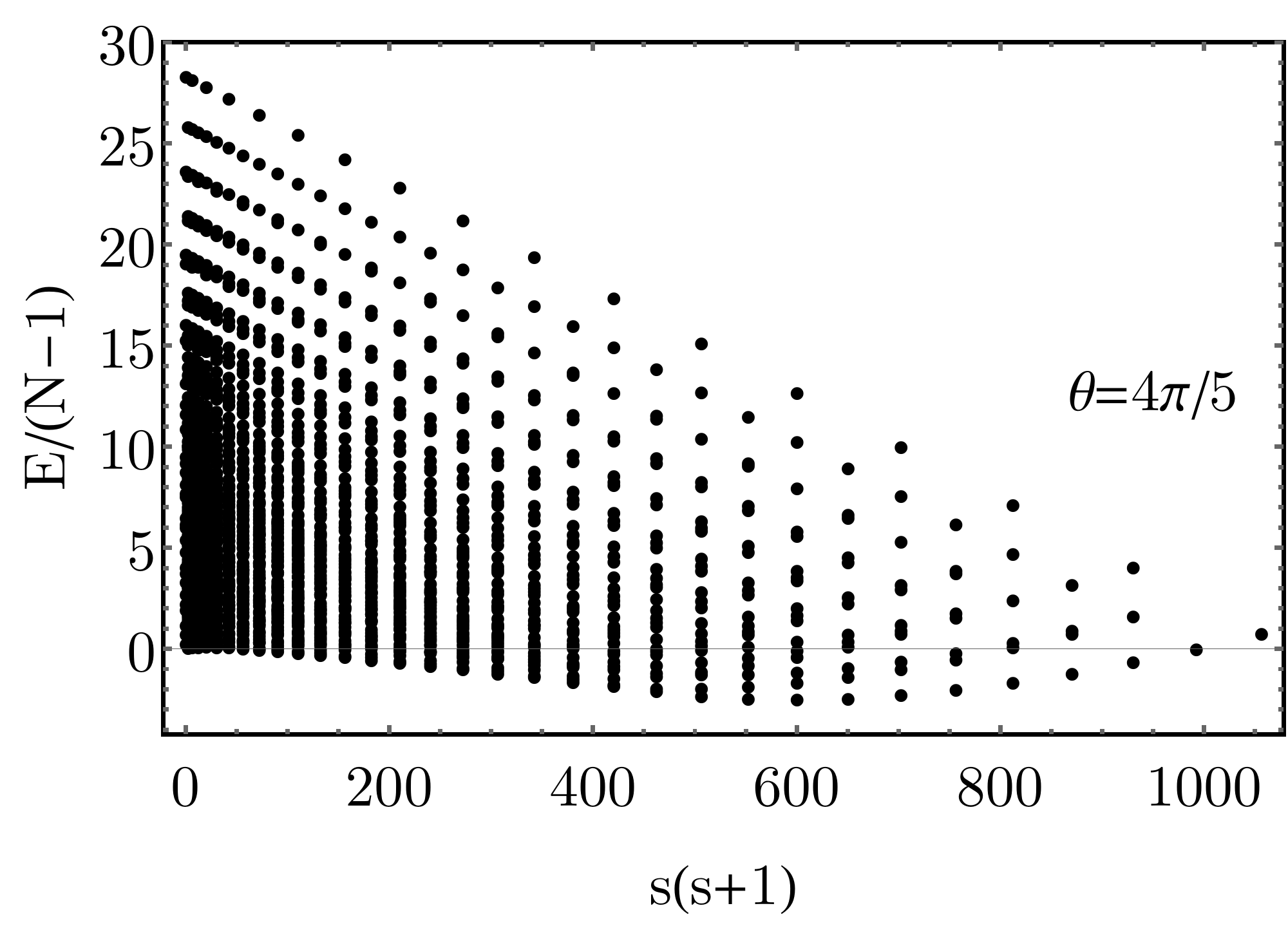}
   \label{fig:sp2}}
 \\
   \subfloat[]{\includegraphics[width=0.5\linewidth]{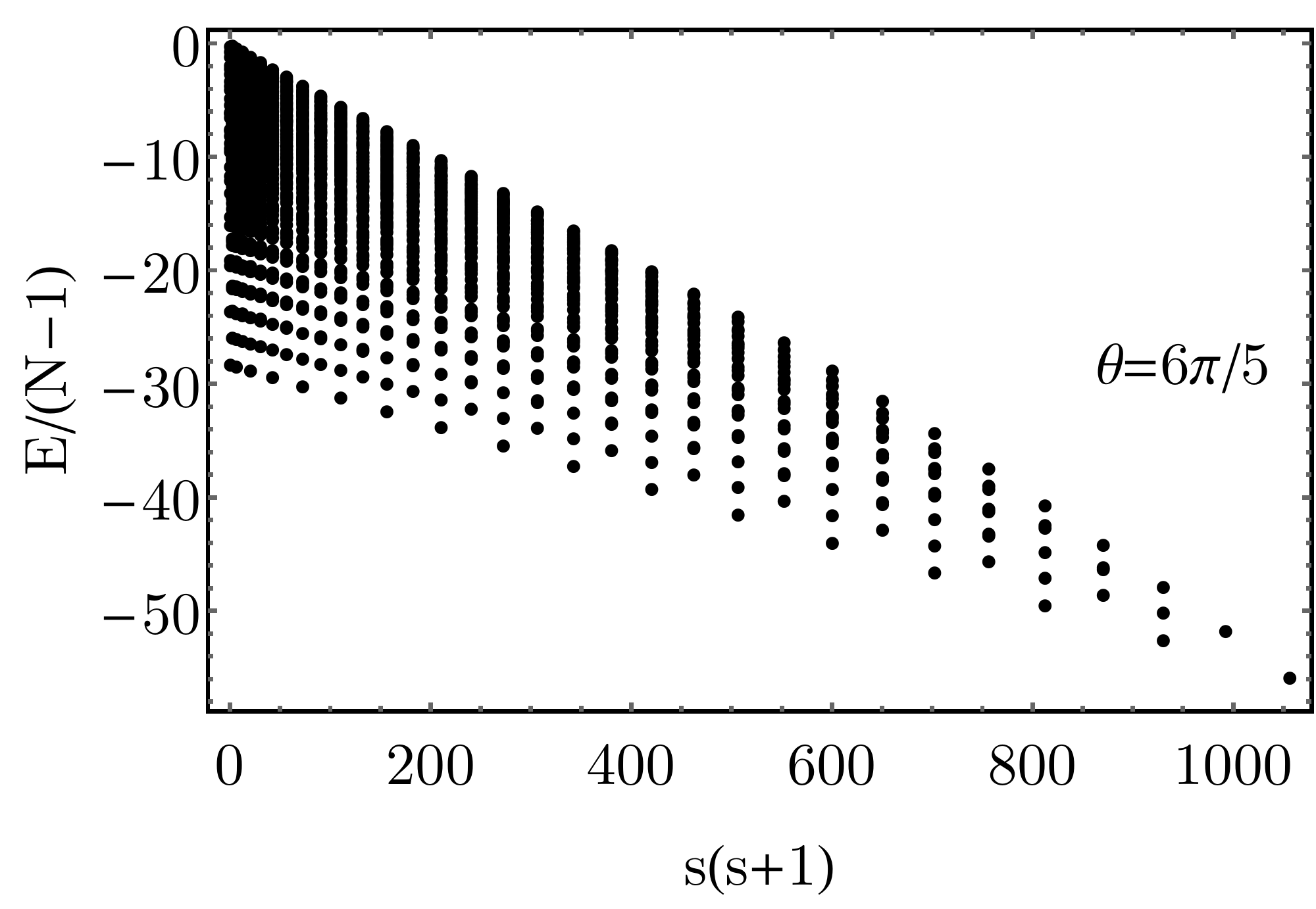}
   \label{fig:sp3}}                                 
   \subfloat[]{\includegraphics[width=0.5\linewidth]{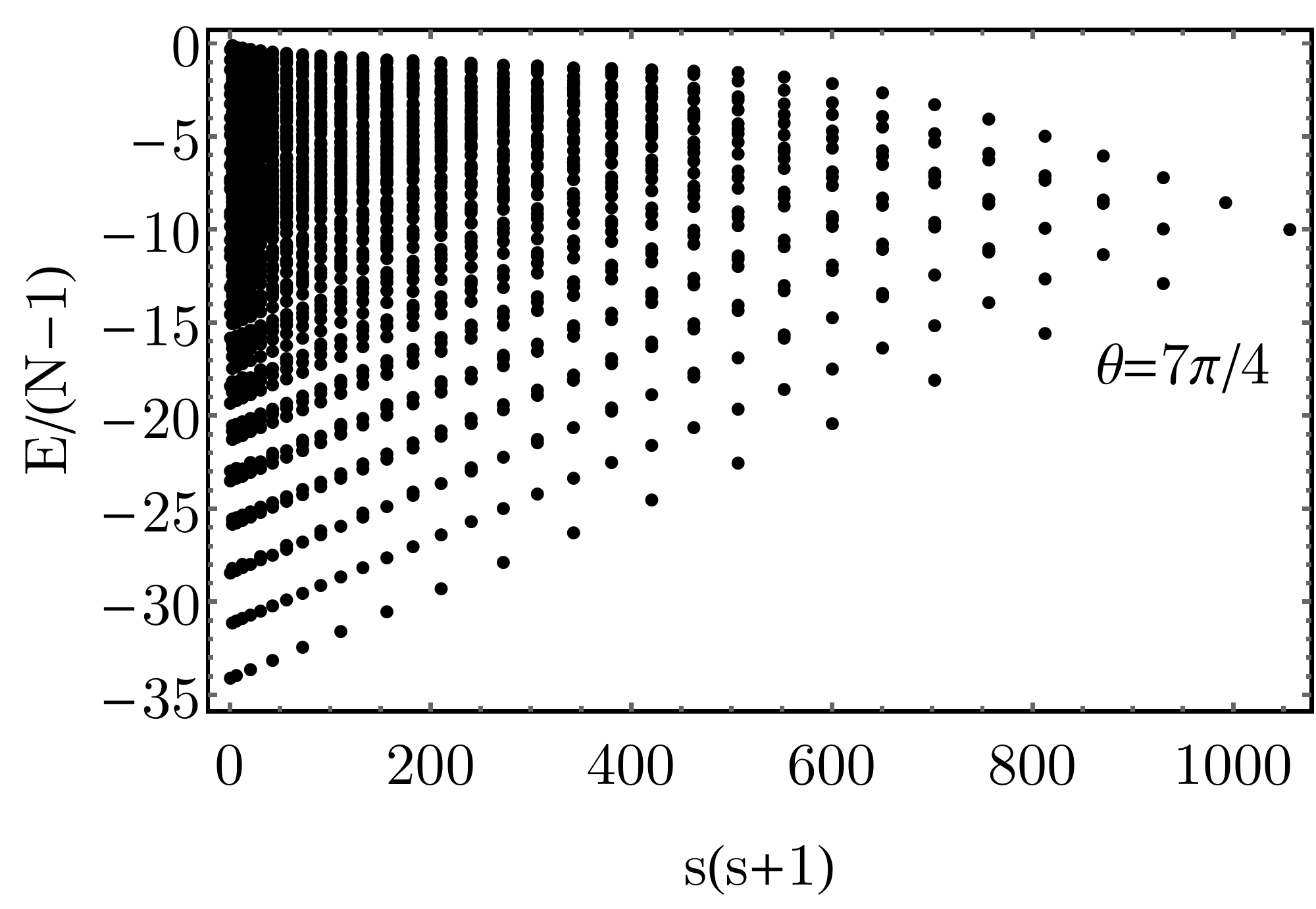}
   \label{fig:sp4}}
 \caption{The energy spectrum of the Hamiltonian \eqref{eq:CH} as a function of the total spin square. Each subfigure represents one of the four ground state phases. The spectrum is evaluated for $N=36$ sites.}
   \label{fig:spectrum}
 \end{figure}
 
Lastly, we discuss the full energy spectrum of our model. 
Let us first note, that as a consequence of the infinite range interaction, the energy given by Eq.~\eqref{eq:CH} is not an extensive quantity. In order to make it extensive in the thermodynamic limit, we normalize our Hamiltonian with an additional $1/(N-1)$ factor. Fig.~\ref{fig:spectrum} illustrates the energy spectrum as a function of $s(s+1)$ for representative values of the $\theta$ parameter for each of the ground-state phases. In Fig.~\ref{fig:sp1}, we have chosen the representative angle $\theta=\pi/4$ from the $\su(3)$ singlet phase. As expected, the lowest energy belongs to $s=0$, and the next level is very close in energy to the ground state. In Fig.~\ref{fig:sp2}, we have chosen $\theta=3\pi/4$ representing the partially magnetized phase. Here, the lowest energy belongs to an intermediate value of the total spin given by $\lambda_1^*(3\pi/4)$. Fig.~\ref{fig:sp3} with $\theta=6\pi/5$ corresponds to the ferromagnetic phase. The lowest energy is for $s=N$. The next lowest energy level is separated by a gap, which does not vanish when $N$ goes to infinity. Finally, Fig.~\ref{fig:sp4} is plotted for $\theta=9\pi/5$ representing the symmetric \su(2) singlet phase. The minimal energy level is at $s=0$ and the separation between the two lowest levels vanishes in the thermodynamic limit.

We further note, that by the transformation $\theta\to\theta+\pi$ the Hamiltonian \eqref{eq:CH} changes sign. 
Therefore, the spectra also get reflected under such a transformation. A reminiscent behavior is approximately present when comparing Figs.~\ref{fig:sp1} and~\ref{fig:sp3}, as well as~\ref{fig:sp2} and~\ref{fig:sp4}, although the four representative $\theta$ angles were intentionally chosen not to be symmetric on the circle.

\begin{figure}[tb!]
   \centering
   \subfloat[]{\includegraphics[width=0.482\linewidth]{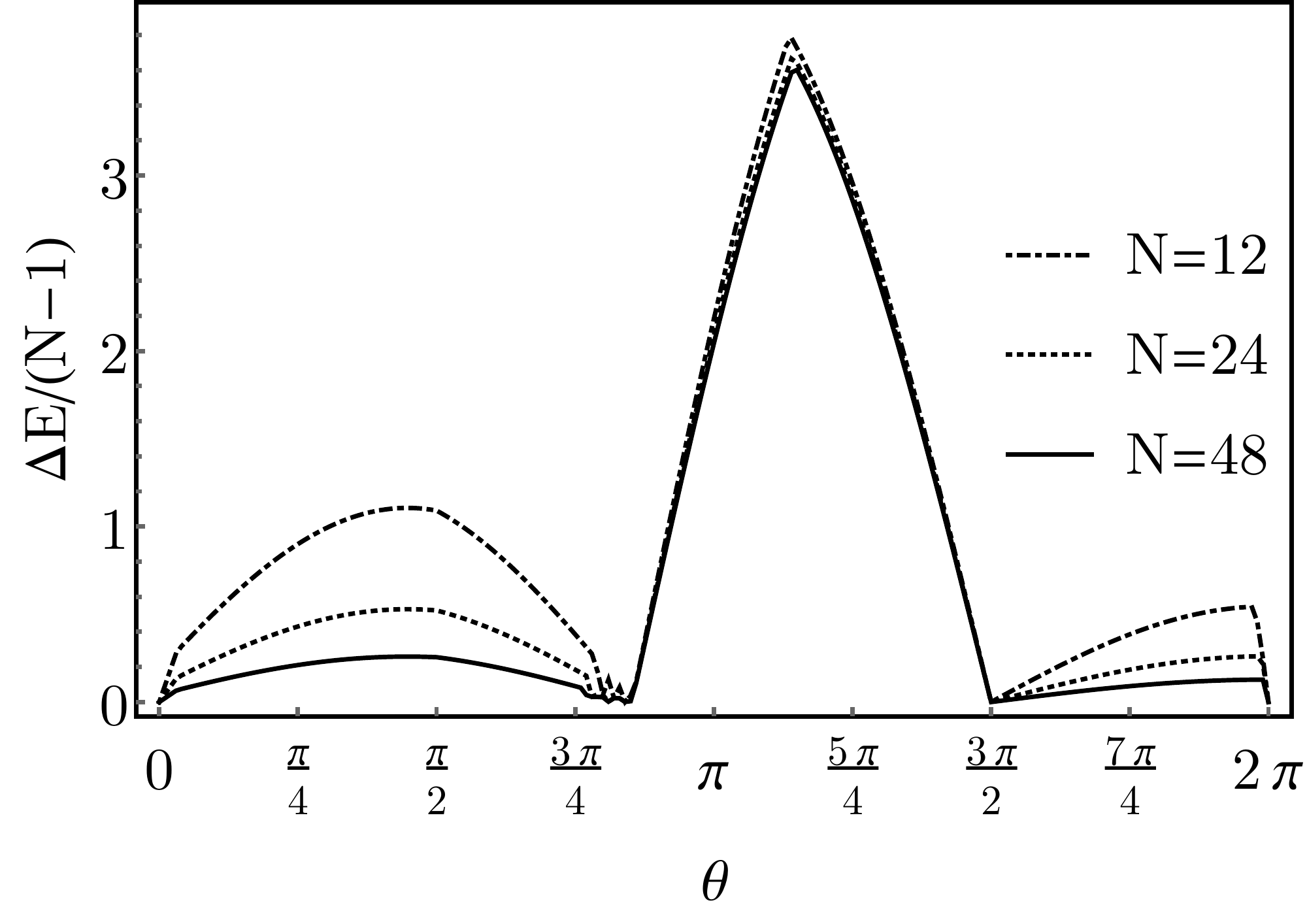}
   \label{fig:gap1}}
   \subfloat[]{\includegraphics[width=0.5\linewidth]{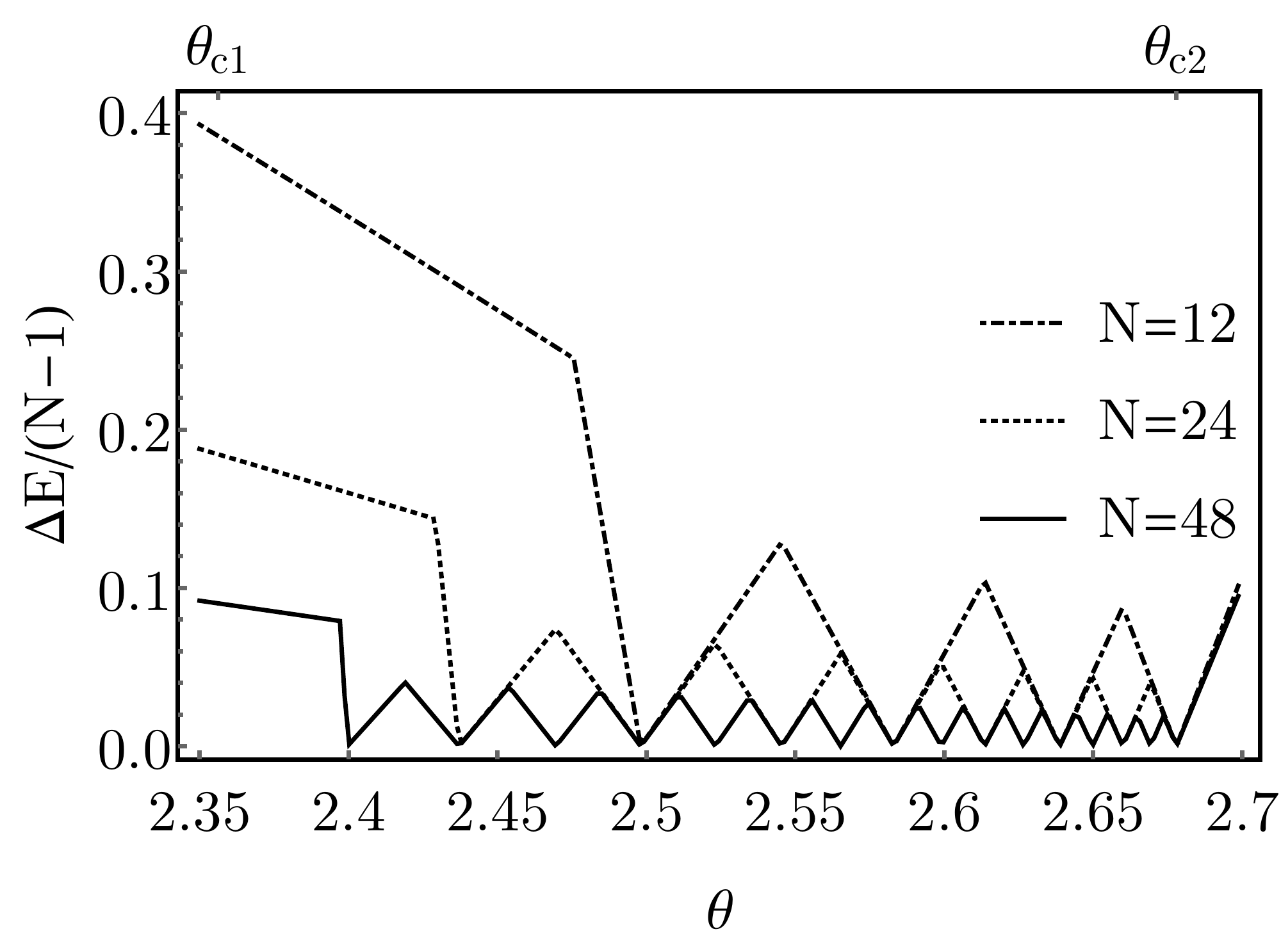}
   \label{fig:gap2}}
 \caption{The difference between the two lowest energy levels of the Hamiltonian \eqref{eq:CH} for $N=36$ sites.}
   \label{fig:gap}
 \end{figure}

Let us take a closer look at the energy gap, i.e., the separation of the two lowest lying energy levels, see Fig.~\ref{fig:gap}. The gap remains finite as $N$ tends to infinity only in the ferromagnetic region. Both in the \su(3) singlet phase and in the symmetric \su(2) singlet phase, the gap tends to $0$ as $\mathcal{O}(1/N)$. The situation is quite delicate in the partially magnetized phase.  For any finite $N$, there is a series of ground-state level crossings, the number of which is proportional to $N$, see Fig.~\ref{fig:gap2}. As $N$ tends to infinity these level-crossings get increasingly dense, and the phase becomes gapless in the thermodynamic limit even without the $1/(N-1)$ normalization of the Hamiltonian.

\section{Summary and outlook}
\label{Sec:discussion} We determined the ground-state phase diagram and energy spectrum of the spin-$1$ bilinear-biquadratic model on the complete graph. In this simple setting, the Hamiltonian is a linear combination of the quadratic Casimir operators of the \su(2) and \su(3) Lie algebras. These two Casimir operators commute, thus the diagonalization of the Hamiltonian reduces to the representation theoretic problem of identifying the embedding of \su(2) irreducible representations into \su(3) ones. Solving this problem is one of the main results of this paper. By diagonalizing the Hamiltonian, we found that the model has four phases belonging to different symmetry sectors. With respect to \su(2) two of the phases are singlets, there is also a partially magnetized and a ferromagnetic phase. One of the \su(2)-singlet phases is also \su(3)-singlet, while the other one belongs to the completely symmetric subspace and is thus characterized by a maximal $\lambda_1=N$ quantum number. The ferromagnetic ground state is also symmetric, i.e., $s=\lambda_1=N$. The most interesting part of our phase diagram is the gapless partially  magnetized phase  between the ferromagnetic and the \su(3)-singlet phase. The $\lambda_1$, $\lambda_2$ and $s$ quantum numbers change gradually within this phase, while they stay constant within the other phases. At all phase boundaries the quantum numbers have a discontinuity proportional to $N$, except at the  boundary between the partially magnetized and the ferromagnetic phase ($\theta_{c2}$).

Spin models on complete graphs are generally believed to mimic the properties of their counterparts on regular lattices in sufficiently high dimensions. However, for the BLBQ model, already some aspects of low-dimensional lattice models are reflected in the complete graph results.
A natural example is the case of ferromagnetic coupling, for which the ground-state space is the $\lambda_1=s=N$ maximally polarized subspace, independently of the underlying geometry. A more interesting observation is that the symmetric \su(2) singlet, $(\lambda_1, \lambda_2)=(N,0), s=0$, appears as a limiting  ground state also in the one-dimensional BLBQ model at
the boundary of the dimerized \su(2)-singlet and the ferromagnetic phase \cite{hu2014berry}. In contrast, dimerization and trimerization, which are important features of the BLBQ-model on bipartite and tripartite lattices, cannot be described on complete graphs. 
Thus, a natural generalization of our problem would be to consider the model on $k$-partite complete graphs.  Already in the bipartite case, one can expect the appearance of symmetry-breaking antiferromagnetism. 

Finally, let us comment about the possible experimental relevance of this model. In Ref.~\cite{beverland2016realizing}, an experiment with ultracold atoms was proposed for the realization of the \SU(3) symmetric point of the BLBQ Hamiltonian on a complete graph. It would be desirable to extend this approach to the entire phase diagram. An experimental realization would not only mean that one can study these quantum phases of matter in the lab, but one would also obtain metrologically useful  states.  In this respect, particularly the singlet phases can be of interest, as macroscopic singlet states have been  proposed for gradient magnetometry \cite{urizar2013macroscopic,apellaniz2017precision}.

\section*{Acknowledgements}

We would like to thank R. Ganesh, Pierre Nataf and  Karlo Penc for useful discussions. This work was supported by the Hungarian  National Research, Development and Innovation Office (NKFIH) through Grant No. K115624, and by the Hungarian Academy of Sciences through Grant No. LP2011-016 and the J\'anos Bolyai Scholarship.

\appendix

\section*{Appendix}

\section{The su(3) Casimir operator in terms of spin operators}
\label{sec:appCasimir}

In this appendix, we express the quadratic Casimir operators of \su(2) and \su(3) with the help of only the \su(2) generators corresponding to the spin-1 representation. Let us first consider the two-site problem. The \su(2) and \su(3) Casimir operators are defined as\footnote{Note that for the \su(2) Casimir, we use the physicist convention which differs by a factor of 2 used in Ref.~\cite{iachello}.}
  \begin{align}
    \label{eq:s1s2}     C^{\su(2)}_2&=\sum\limits_{\mu=1}^3[S_1^{\mu}S_1^{\mu}+S_2^{\mu}S_2^{\mu}+2S_1^{\mu}S_2^{\mu}]=4\idm+\sum\limits_{\mu=1}^32S_1^{\mu}S_2^{\mu},\\
 C^{\su(3)}_2&=\sum\limits_{\alpha=1}^8[\Lambda_1^{\alpha}\Lambda_1^{\alpha}+\Lambda_2^{\alpha}\Lambda_2^{\alpha}+2\Lambda_1^{\alpha}\Lambda_2^{\alpha}]= \frac{32}{3}\idm+\sum\limits_{\alpha=1}^82\Lambda_1^{\alpha}\Lambda_2^{\alpha}.
  \end{align}
In the two-site tensor product space, let us introduce the projections to the spin-$s$ irreducible subspaces, denoted by $P_s$, with $s\in\{0,1,2\}$.
Obviously, $\idm=P_0+P_1+P_2$, and $C_2^{\su(2)}=2P_1+6P_2$. Furthermore, using \eqref{eq:s1s2} we can also express $S_1^\alpha S_2^\alpha$ as,
\begin{align}
\sum\limits_{\mu=1}^3S_1^{\mu}S_2^{\mu}&=-2P_0-P_1+P_2,\\
\left(\sum\limits_{\mu=1}^3S_1^{\mu}S_2^{\mu}\right)^2&=4P_0+P_1+P_2.
\end{align}
By inverting these relations, we obtain
\begin{subequations}
\label{eq:projectors}
\begin{align}
P_0&= -\frac{1}{3} \idm + \frac{1}{3} \left(\sum\limits_{\mu=1}^3S_1^{\mu}S_2^{\mu}\right)^2, \\
P_1&= \idm -\frac{1}{2} \sum\limits_{\mu=1}^3S_1^{\mu}S_2^{\mu} -\frac{1}{2} \left(\sum\limits_{\mu=1}^3S_1^{\mu}S_2^{\mu}\right)^2,\\
P_2&= \frac{1}{3} \idm + \frac{1}{2} \sum\limits_{\mu=1}^3S_1^{\mu}S_2^{\mu} + \frac{1}{6} \left(\sum\limits_{\mu=1}^3S_1^{\mu}S_2^{\mu}\right)^2.
\end{align}
\end{subequations}

The  the two-site tensor product space $\mathbb{C}^3\otimes\mathbb{C}^3$ decomposes into a direct sum of two irreducible \su(3) subspaces, namely the symmetric [$(\lambda_1,\lambda_2)=(2,0)$] and the antisymmetric [$(\lambda_1,\lambda_2)=(1,1)$] one. Thus, the  Casimir $C_2^{su(3)}$ is a linear combination of the projection operators corresponding to these subspaces, $C_2^{su(3)}= \tfrac{40}{3} P_{(2,0)} + \tfrac{16}{3} P_{(1,1)}$.
Since \su(2) is a subalgebra of \su(3), each \su(3) irreducible subspace is a direct sum of \su(2) irreducible subspaces. Therefore, the projections to the \su(3) irreducible subspaces are sums of the corresponding \su(2) projections,
\begin{equation}
  P_{(2,0)}=P_0+P_2\, , \quad P_{(1,1)}=P_1 \,.
\end{equation}
Utilizing this and Eqs.~\eqref{eq:projectors}, we can now express $C_2^{SU(3)}$ in the desired form:
\begin{equation}
  C_2^{SU(3)}=\frac{40}{3}P_0+\frac{16}{3}P_1+\frac{40}{3}P_2=\frac{16}{3}\idm+4\sum\limits_{\mu=1}^3S_1^{\mu}S_2^{\mu}+4(\sum\limits_{\mu=1}^3S_1^{\mu}S_2^{\mu})^2. 
\end{equation}
As $ C_2^{SU(3)}= \tfrac{32}{3}\idm + 2 \sum\limits_{\alpha=1}^8\Lambda^{\alpha}_1\Lambda^{\alpha}_2$, we are now able to express $\sum\limits_{\alpha=1}^8\Lambda^{\alpha}_k\Lambda^{\alpha}_l$ in Eq.~\eqref{eq:su3casexp} with spin operators:
\begin{equation}
  \sum\limits_{\alpha=1}^8\Lambda^{\alpha}_k\Lambda^{\alpha}_l=-\frac{8}{3}\idm+2\left[\sum\limits_{\mu=1}^3S_k^{\mu}S_l^{\mu}+(\sum\limits_{\mu=1}^3S_k^{\mu}S_l^{\mu})^2 \right].
\end{equation}
Finally, we can express the N-site Casimir, and arrive to the result expressed in Eq.~\eqref{eq:su3casexp}.

\section{Schur-Weyl duality}
\label{SchurWeylApp}

Here we recall some important results about associating the irreps  of $\SU(d)$ and its Lie algebra $\su(d)$  with the irreps of the permutation group $S_N$, labeled by the Young diagrams. For this end, let us first discuss the Schur-Weyl duality, which provides a one-to-one mapping between the Young diagrams of at most $d$-rows and the irreps of \U(d).
The matrices corresponding to the $N$-fold tensor product representation of $\U(d)$ commute  with the natural action of $S_N$ on $(\CC^d)^{\otimes N}$. The Schur--Weyl duality states that under the joint action of these two groups the Hilbert space decomposes as
\begin{equation}
  (\CC^d)^{\otimes N}= \bigoplus_\lambda \kk_\lambda \otimes \HH_\lambda,
  \label{SW}
\end{equation}
where $S_N$ acts trivially on each $\HH_\lambda$ and irreducibly on each $\kk_\lambda$, while $\U(d)$ acts irreducibly on each $\HH_\lambda$ and trivially on each $\kk_\lambda$. The direct sum in Eq.~\eqref{SW} goes over all $N$-box Young diagrams with at most $d$ rows, or alternatively all $d$-tuples $(\lambda_1,\cdots,\lambda_d)$ of integers satisfying the conditions $\lambda_1+\lambda_2+\cdots+\lambda_d=N$ and $\lambda_1\ge\lambda_2\ge\cdots\ge \lambda_d\ge 0$. 
The dimension of $\kk_\lambda$, i.e.,  the dimension of the $S_N$ irrep corresponding to the Young diagram $\lambda$, is given by the hook length formula \cite{sagan2013symmetric}:
\begin{equation}
  \label{eq:hook}
 m_\lambda=\text{dim}(\kk_\lambda)=\frac{N!}{\prod\limits_{i,j}h_\lambda(i,j)}. 
\end{equation}
Here $h_\lambda(i,j)$ is the number of boxes in the ``hook'' at the $i\ts{th}$ row and $j\ts{th}$ column of the diagram $\lambda$, meaning the number of boxes at positions $(k,l)$ such that $i=k$ and $l\ge j$ or $i\ge k$ and $l=j$. For each irrep of $\U(d)$, there exists a unique $N$ for which the irrep appears in the decomposition \eqref{SW}\cite{hamermesh2012group}. 
Moreover, if $\lambda$ and $\lambda'$ are Young diagrams appearing in the direct sum in Eq.~\eqref{SW} for which $\lambda\neq \lambda'$, then the $\U(d)$ irreps acting on $\HH_{\lambda}$ and $\HH_{\lambda'}$ are inequivalent. Thus, the Schur--Weyl duality provides a one-to-one mapping between $\U(d)$ irreps and Young diagrams with at most $d$ rows.

Associating Young diagrams with the irreps of \SU(d) is slightly more complicated. First, let us note that an irrep of \U(d) restricted to the \SU(d) subgroup remains irreducible, moreover, all \SU(d) irreps arise this way. However, inequivalent \U(d) irreps may provide equivalent \SU(d) irreps. More explicitly, two \U(d) irreps with Young diagrams $\lambda$ and $\lambda'$ reduce to the same \SU(d) irrep if and only if
\begin{equation}
\label{eq:equivalencerelation}
\lambda_1-\lambda_1'=\lambda_2-\lambda_2'=\cdots=\lambda_d-\lambda_d'.
\end{equation}
In other words, each $\SU(d)$ irrep is labeled uniquely by an equivalence class of Young diagrams which are related to each other by attaching or cutting off columns of height $d$ from the left-hand side. In every equivalence class there is one diagram with $\lambda_d=0$, so we are effectively labeling the $\SU(d)$ irreps with diagrams of at most $d-1$ rows. Applying this labeling for the \SU(d)-irrep decomposition of $(\CC^d)^{\otimes N}$, instead of all $N$-box diagrams with at most $d$ rows, we get all diagrams with $N-i\cdot d$ boxes, where  $i=0,1,\cdots \lfloor N/d \rfloor$, and at most $d-1$ rows.  In particular, for the \SU(3) case, all the irreps can be uniquely labeled by 2-row Young diagrams, see Fig.~\ref{fig:possiblel1l2}.
\bibliography{blbq}

\end{document}